\documentclass[12pt]{article} % it's 12pt

\usepackage[hidelinks]{hyperref}
\hypersetup{allcolors=blueish,colorlinks=true}
\usepackage{scicite}

\usepackage{times}

\usepackage{amsmath}
\usepackage{amsfonts}
\usepackage{amssymb}
\usepackage{graphicx}

\usepackage{booktabs}

\newenvironment{sciabstract}{%
\begin{quote} \bf}
{\end{quote}}

\usepackage[labelfont=bf,figurename=Fig.,labelsep=period,font=footnotesize]{caption}

\title{\Large\bf Browsing behavior exposes identities on the Web}

\author
{
Marcos Oliveira${}^{1,\ast}$, Junran Yang${}^{2}$, Daniel Griffiths${}^{1}$, \\Denis Bonnay${}^{3,\dagger}$, and Juhi Kulshrestha${}^{2,\ddagger}$\vspace{.1in}\\
\small{${}^{1}$University of Exeter, Exeter, UK}\\
\small{${}^{2}$Aalto University, Espoo, Finland}\\
\small{${}^{3}$Université Paris Nanterre, Paris, France}\vspace{.1in}\\
\footnotesize{
$^\ast$\href{mailto:moliveira@tuta.io}{moliveira@tuta.io}
$^\dagger$\href{mailto:denis.bonnay@gmail.com}{denis.bonnay@gmail.com} 
$^\ddagger$\href{mailto:juhi.kulshrestha@aalto.fi}{juhi.kulshrestha@aalto.fi}}\vspace{-.4cm}
}
\date{}

\usepackage{changepage}

\usepackage{comment}

\usepackage{xcolor}
\usepackage{soul}
\definecolor{reddish}{HTML}{FBB4AE}
\definecolor{blueish}{HTML}{1F54A9}
\definecolor{magentish}{HTML}{FF00AA}
\definecolor{greenish}{HTML}{a1d99b}

\usepackage{setspace}

\usepackage{amsmath}           % For the use of math

\usepackage{lineno}
\usepackage{mdframed}
\usepackage{wrapfig}

\begin{document} 



\maketitle 

% \newpage 


% The abstract should be a single paragraph, not to exceed 250 words and ideally closer to 200, written in plain language that a general reader can understand. It should include
% An opening sentence that states the question/problem addressed by the research AND
% Enough background content to give context to the study AND
% A brief statement of primary results AND
% A short concluding sentence.
% Do not include citations or undefined abbreviations in the abstract. Any abbreviations that appear in the title should be defined in the abstract.

% \baselineskip18pt % "one" spacing

\begin{sciabstract}
Abstract

\textnormal{How easy is it to uniquely identify a person based solely on their web browsing behavior? Here we show that when people navigate the Web, their online traces produce fingerprints that identify them. Merely the four most visited web domains are enough to identify 95\% of the individuals. These digital fingerprints are stable and render high re-identifiability. We demonstrate that we can re-identify 80\% of the individuals in separate time slices of data. Such a privacy threat persists even with limited information about individuals' browsing behavior, reinforcing existing concerns around online privacy.  
}
\end{sciabstract}

% \red{GUYS, USE THE OTHER TEMPLATE; THIS one is FOR ARXIV LATER ON...}

% \baselineskip15pt % "one" spacing
\baselineskip20pt 

\section*{Introduction}
In the era of ubiquitous technology, our online habits have become a goldmine for companies seeking to extract value from our data, often risking our privacy~\cite{zuboff_surveillance_2019}. By looking at how we browse the Web, they learn about us, enabling them to build highly targeted advertising campaigns and monetize user profiles with third-party advertisers~\cite{white_social_2020,cotter_reach_2021}. Such lucrative business models focus on tracking, understanding, and predicting individuals' behavior~\cite{zuboff_surveillance_2019}. However, despite the commercial success, the behavioral traits that enable this profitability remain elusive and quantitatively unexplored. 

Bridging the gap between web browsing data and behavioral science, researchers have recently shown that users' online behavior is highly predictable, revealing an average of 85\% potential predictability~\cite{kulshrestha_web_2021}. Much of this predictability arises from users' online habits. Despite the myriad options available online, individuals adhere to web routines (i.e., repeated browsing patterns or habitual website visits) that reduce the uncertainty about their behavior~\cite{kulshrestha_web_2021,schnauber2023routines,piccardi2023large,piccardi2023curious}. This innate human proclivity for habits extends to offline activities such as shopping and mobility, rendering them equally predictable~\cite{song2010limits,krumme_predictability_2013,barbosa2018,pacheco2022predictability}. Such high predictability enables businesses to tailor individual-level services and products, forming a cornerstone of the contemporary data economy.

While this predictability helps refine user experience by improving personalization and advertising relevance, it raises concerns about individuals' privacy and autonomy~\cite{susser_technology_2019}. By understanding and predicting users' behavior, platforms can exploit psychological vulnerabilities, drawing attention to a core worry of surveillance capitalism: predicting user behavior to modify it~\cite{zuboff_surveillance_2019,susser_technology_2019}. This paradigm touches on various facets of an individual's life, such as health, well-being, consumption, and voting patterns~\cite{blease_open_2023,cosgrove_psychology_2020,ford_hormonal_2021,chester_digital_2019}. To enable this practice, business models rely on the precise identification of users to target them effectively.

In diverse contexts such as shopping and mobility, this identification is facilitated by the uniqueness of individuals' actions serving as fingerprints~\cite{de_montjoye_unique_2013,de_montjoye_unique_2015}. For instance, how individuals navigate within cities produces distinct signatures: merely four spatiotemporal points can pinpoint most users in mobile phone data sets containing time-location details~\cite{de_montjoye_unique_2013}. Likewise, minimal data on purchasing activities can re-identify individuals within anonymized credit card records~\cite{de_montjoye_unique_2015}. Such a distinctiveness in the way people behave allows behavioral traits to identify individuals, replacing traditional personally identifiable information (e.g., email address, phone number).
\emph{While our habits make us predictable, our uniqueness makes us identifiable.}

% \red{POTENTIAL PLACE TO MAKE THE CONCEPT OF UNIQUENESS MORE TANGIBLE (Q1)}
% Humans, however, are unique in their actions, facilitating identification. 

Previous works have explored similar uniqueness in online environments from a technical standpoint, shedding light on browser fingerprinting~\cite{gomez-boix_hiding_2018} and security vulnerabilities~\cite{olejnik_why_2012,bird_replication_2020}. For example, inadvertent leaks of browsing history can identify users. Specifically, listing which of the top 50 globally most-visited websites a user visited is an efficient fingerprint~\cite{olejnik_why_2012}. This strategy uniquely identifies individuals in a data set containing whether users have once visited websites from a predefined list. Similarly, unique preferences in movies~\cite{narayanan_robust_2008}, Facebook interests~\cite{gonzalez-cabanas_unique_2021}, and online sharing patterns~\cite{su_-anonymizing_2017} can compromise anonymity within respective data sets.
However, while previous works have characterized the specifics of users' preferences among a fixed set of options, they have overlooked the intrinsic uniqueness embedded in users' habits in the wild~\cite{kulshrestha_web_2021}.

In this work, we uncover how our browsing habits uniquely identify us. By examining the complete data of users' browsing behavior, we unravel the identifying power of their habitual websites. These consistently visited web domains serve as a fingerprint. We demonstrate that by knowing just the top four most-visited domains of a user, we can unmask 95\% of users in our data set. This staggering rate of identifiability highlights a vulnerability within our digital routines: our habits are distinctive, rendering us uniquely identifiable. 
This identification precision remains even with data from only a limited set of domains (i.e., sparse tracking).
Even more alarmingly, when analyzing the data across varied time frames, we can re-identify 80\% of users across different intervals. Such an inherent risk of de-anonymization due to our habits reinforces the concerns surrounding online privacy, emphasizing the urgency to protect our digital footprints.

% \vspace{1em}
% \begin{mdframed}[
%         backgroundcolor=black!4,
%         bottomline=false,
%         rightline=false,
%         topline=false,
%         leftline=false,
%         splittopskip=1mm,%
%         splitbottomskip=1mm,%
%         skipabove=1em,%
%         skipbelow=1em,%
%         leftmargin=0cm,%
%         rightmargin=0cm,%
%         innertopmargin=1em,%
%         innerrightmargin=1em,%
%         innerleftmargin=1em,%
%         innerbottommargin=1em,%
%     ]{\small\noindent \textbf{Significance Statement.}
%     In the digital age, seemingly benign data, like %such as 
%     our most visited websites, can compromise privacy. We show that with minimal data, 95\% of individuals can be uniquely identified with web tracking data, and 89\% can be re-identified over different periods. Our findings reveal a substantial loss of digital anonymity and uncover a fundamental aspect of privacy: the very nature of our online behavior makes us both predictable and identifiable, underscoring the need to understand and safeguard our digital footprints. }
% \end{mdframed}

\section*{Results} 
We analyze users' web browsing behavior by examining web tracking data from 2,148 users in Germany, which contains 9,151,243 user visits to websites, including visit time and active seconds spent on the page, collectively spanning 49,918 unique domains (see Methods for details).

\subsection*{From web browsing traces to web fingerprints}
\begin{figure*}[b!]
\centering
\begin{adjustwidth}{-.3in}{0in} 
\includegraphics[width=\linewidth]{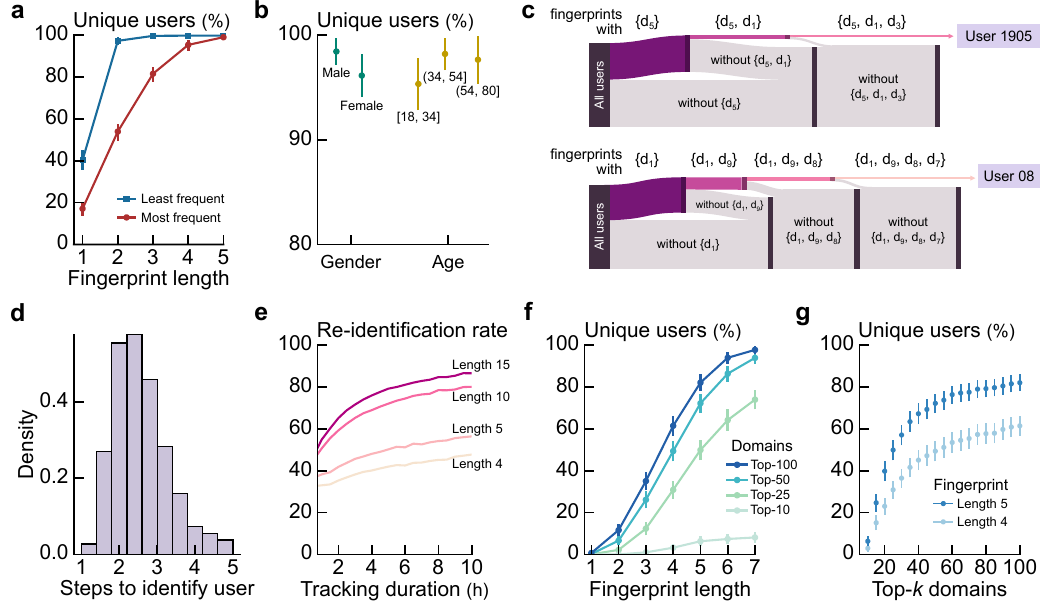}
\end{adjustwidth}
% \begin{adjustwidth}{-4em}{+4in} 
\captionsetup{width=1.15\linewidth}
\caption{\textbf{People’s habitually visited websites serve as fingerprints that distinguish them.} 
%
% \includegraphics[width=6.95in]{figures/Panel_1_v5.pdf}
% \caption{\textbf{People’s habitual websites serve as fingerprints that distinguish them apart.} 
%
%
\textbf{(a)}~The percentage of users with unique most-visited domains list (i.e., fingerprint) with varying fingerprint length. 
Almost all users have unique four-domain fingerprints (i.e., four most visited domains). 
\textbf{(b)}~The percentage of users with unique four-domain fingerprints grouped by age and gender. 
Regardless of gender and age, a four-domain fingerprint yields high uniqueness.
% with males marginally higher at 98\% compared to females at 96\%, and age groups 34-54 and 55-80 achieving 98\% and 97\% uniqueness, slightly above the 18-34 group's 95\%. 
%
\textbf{(c)} A schematic of a step-by-step user identification via users' habitual websites. 
For each user, we randomly select a domain from their list of five most-visited domains then group all users sharing the same domain. By selecting additional domains, this process progressively refine groups, until the user is uniquely isolated. In the illustration, bar sizes represent proportional number of users.
\textbf{(d)}~The distribution of steps to identify a user within our data. 
While four domains ensure uniqueness, fewer domains are often enough for identification;
by following steps in \textbf{(c)}, we need
an average of $2.45$ steps to identify users.
\textbf{(e)}~The re-identification rate for different duration of data collection. 
The fingerprints enable re-identification of most individuals in separate time slices of data. 
% The longer the browsing data collected, the higher the re-identification rates. 
%
\textbf{(f)}~The percentage of unique users in scenarios of fewer tracking domains. 
By collecting data of a limited number of domains only, the majority of users can still be distinguished. 
% The rate of unique users reaches 82\% with just the Top-100 domains, using a fingerprint length of 5.
%
\textbf{(g)}~The percentage of unique users in scenarios of fewer tracking domains, with fixed fingerprint lengths. Collecting data from more domains yields higher uniqueness, but with decreasing returns. 
% 
% , and gray areas indicate users who have been filtered out.
}
% textwidth: \printinunitsof{in}\prntlen{\textwidth}
% linewidth: \printinunitsof{in}\prntlen{\linewidth}
% textwidth: 6.99866in linewidth: 6.99866in
\label{fig1}
\end{figure*}

To understand how online browsing behavior yields fingerprints on the Web, we analyze individuals' most visited web domains. We demonstrate that we can uniquely identify an individual by using their list of most frequented domains. First, we describe each individual with an $n$-tuple comprising their $n$ most visited domains; then, we count the number of non-duplicate $n$-tuples within our data, thereby identifying the percentage of unique users (i.e.,~unique fingerprints; see Fig.~\ref{fig1}a). Our results reveal that 95\% of the users have unique fingerprints when $n=4$, implying that an individual's four most visited domains uniquely identify them. 
 % 0.9530598262183444
 
 This short-length fingerprint identifies users irrespective of gender and age. By grouping users based on demographics and comparing their four-domain fingerprints, we find a consistent high proportion of unique fingerprints with only minimal group variations. Our results show a marginally higher percentage of unique male individuals at 98\%, compared to 96\% for female individuals (Fig.~\ref{fig1}b). Likewise, unique users are prevalent across age groups: 98\% and 97\% for the 34-54 and 55-80 brackets, with the 18-34 group slightly lower at 95\%. Furthermore, we also examine group variations in fingerprint lengths, finding no statistical differences when accounting for individuals' demographics (see SI). 
% \red{add CI}
% (male, female)
% (0.9843019746012931, 0.9613283755924689)
% (18-34, 34-54, 55-80)
% (0.9533322944262638, 0.981995250036487, 0.9765014060634076)

Though most users are unique in their four most-visited domains, we often need fewer data points for \textit{user identification}. To determine how many domains are needed to pinpoint a user, we examine fingerprints at the individual level. For each unique user~$i$, we randomly select a domain from their fingerprint and group all users who have that domain in their fingerprints (see Methods). Then, we select another most-visited domain from user $i$ and narrow our group to those with both domains~(Fig.~\ref{fig1}c). We repeat this step, incrementally adding domains, until we isolate user $i$. At this point, we have a set of domains which exists only within user $i$'s fingerprint. Our analysis shows that we need an average of $2.45$ steps to identify a unique user within the data set (Fig.~\ref{fig1}d). This finding indicates that although four domains guarantee uniqueness, users' distinct online habits facilitate their identification with fewer domains.

We note that fingerprints would be shorter if it were not for popular domains---more than 80\% of the users share their most-visited domain with other users (Fig.~\ref{fig1}a). These popular domains enlarge users' fingerprints, whereas unpopular domains make them distinctive. For example, when we use the least visited domains to form the $n$-tuples, we find that most individuals have unique fingerprints when $n\geq2$, implying that the two least visited domains of a user uniquely identify them (Fig.~\ref{fig1}a). However, we expect considerable fluctuations in the list of the least visited domains over time, given that these may represent sporadic visits to isolated domains, which raises the question of actual identifiability based on these fingerprints.

\subsection*{Using fingerprints to re-identify individuals}
To explore the fingerprints' efficacy in identifying individuals on the Web, we examine time slices of data to quantify re-identification rate of user fingerprints. We demonstrate that fingerprints enable us to re-identify most individuals in separate time slices of data. First, we partition the data into two time slices, producing two contiguous separate data sets. For each one, we create users' fingerprints independently. Second, we use the fingerprints from the first data set to identify users in the second set and assess the accuracy of this re-identification. We split data based on user browsing time and find an 60\% success rate in re-identifying users using only ten hours of browsing data with a fingerprint of length $n = 5$, reaching success rate of 80\% and 90\% when $n = 10$ and $n = 15$, respectively (Fig.~\ref{fig1}e).

Our results also reveal that collecting more user browsing data improves re-identification rates but with diminishing gains. For example, with a fingerprint of $n = 15$, collecting two hours of users' browsing activity yields a re-identification rate of 65\%. By extending this data collection for two hours, this re-identification raises to 76\%. However, after 6 hours of data collection, the gain in re-identification rate is around 2\% per additional hour. Despite these diminishing returns, our findings indicate that short-duration tracking periods can be efficient for user identification. While this analysis explored the impact of tracking depth, we next examine the effect of tracking breadth.

\subsection*{Using limited knowledge to identify users} 
We investigate the number of domains necessary for user identification, revealing that data from a limited set of domains is sufficient to identify users. First, we compile an ordered list of the most frequently visited domains by all users within our data. For each user, given a specific $k$, we only consider their visits to the Top-$k$ domains in that list, effectively filtering out all other domain visits. Then, we examine the number of unique users based on fingerprints using this limited browsing data. Our analysis reveals that using data exclusively from the Top-100 domains yields unique fingerprints for 82\% of users when $n=5$ (Fig.~\ref{fig1}f). This result implies that tracking users via just 0.2\% of all domains would render most users identifiable.

This uniqueness rate increases with the number of domains but with decreasing returns. We find that beyond a certain point, considering more domains only generates marginal growth in unique users. For example, while expanding the domain count from Top-25 to Top-50 boosts uniqueness rates from 50\% to 75\%, further expansion to Top-100 yields a relatively smaller increase to 80\%. To understand this contrast, we focus on fingerprint length $n=4$ and $n=5$ while varying $k$ (Fig.~\ref{fig1}g). We observe that the percentage of unique users rises steeply with $k$, a growth that peaks at around $k=50$; after this point, the growth rate reduces, implying that additional domains contribute only incrementally to user uniqueness.

\section*{Discussion}
In a world where internet usage surpasses 5 billion people, our online habits are a valuable asset for business models relying on tracking and predicting individuals' behavior~\cite{we_are_social,zuboff_surveillance_2019}. However, the extent to which online behavioral traits can be exploited has remained largely unquantified. In this work, we show that online habits serve as fingerprints that identify individuals on the Web. Previous works have explored unique users' preferences but have missed the uniqueness of users' habitual websites. We demonstrate that these consistently visited domains are unique among users, revealing that individuals have distinctive habits that render them uniquely identifiable. 
 % \red{PUNCHINESS (Q2)}  among users
 % % Our results indicate that these consistently visited domains are unique among users, acting as an efficient tool for identification. \red{POTENTIAL PLACE TO MAKE THE CONCEPT OF UNIQUENESS MORE TANGIBLE (Q1)?}
% Such a distinctiveness in the way people behave allows behavioral traits to identify individuals, replacing traditional personally identifiable information (e.g., email address, phone number).
% This staggering rate of identifiability highlights the vulnerability within our digital routines: despite a vast online world, our distinctive habits render us uniquely identifiable. 
% We demonstrate that the distinctiveness in the way people behave on the Web XXXXX identify individuals; the vulnerability within our digital routines: despite a vast online world, our distinctive habits render us uniquely identifiable. 

Our research rests on an important distinction between identifiability and re-identifiability. At a given moment in time, we say that data makes an individual \textit{identifiable} if its restriction to the person is unique to that person. By contrast, we say that data makes an individual \textit{re-identifiable} across two different time slices if its restriction to the person is the same for those two different time slices. In the case of browsing data, we have shown that high identifiability also goes with high-reidentifiability for adjacent time slices. While such adjacent time slices are relevant in some real-life situations (e.g. pushing ads to a targeted individual within a narrow time frame of actions), it would be interesting to look into more long term re-identifiability properties. Biological fingerprints do evolve, but only slightly: what about digital fingerprints?

At any rate, 
this privacy threat is critical given the growing concern among users about their digital \mbox{footprints---a} recent survey found that seven out of every ten people have taken actions to protect their online identity, from disabling cookies to using virtual private networks~\cite{norton_survey}. Despite such preventive actions, our work uncovers a privacy threat transcending technology, rooted in the very nature of our online browsing behavior. We tend to visit the same websites, producing consistent digital fingerprints that can potentially expose our identity. Such a risk is a reflection of our habits rather than technological artifacts. % this could be a place to include \red{Q5 cookies controversies}

Still, the online ecosystem amplifies this risk:~user data collection is prevalent and concentrated in the hands of a few key players~\cite{schelter_ubiquity_2018}. For example, Google AdSense scripts are embedded in over 51 million websites, including 20\% of the top 1 million most popular sites~\cite{adsense_usage}. Such a pervasive data collection enables extensive user monitoring by a single company, which is alarming given our findings that data from only a few domains is enough to identify most users. 

In summary, our work underscores a crucial yet overlooked aspect of online privacy: habitual web visits can unwittingly compromise our digital anonymity. These browsing regularities, a natural by-product of how we browse the Web, pose a substantial privacy risk, challenging efforts to protect online identities. As the online landscape becomes more dominated by extensive data collection, our findings reinforce the urgent need to deal with privacy concerns related to all of our digital traces and not just so-called personal identifiable information, calling for innovative measures to safeguard our digital footprints.

\section*{Methods}

\subsection*{Data}
Our data set consists of the web activity of 2,148 German users in October 2018~\cite{webtrackingdata}. It includes each user's visits to websites, containing anonymized URLs, the corresponding non-anonymized domains, the dates and times of visits, and the time spent on each visit. In total, these users visited 9,151,243 URLs across 49,918 unique domains. In addition, we have each user's self-reported gender and age from survey data; in terms of demographics representation, the data is a representative sample, with respect to gender and age, of German internet users under 65 years old~\cite{kulshrestha_web_2021}.

\subsection*{Uniqueness, identification, and re-identification}
For each user $u$, we define an $n$-tuple $m^n_u = (d^u_1, d^u_2, \cdots, d^u_n)$ as the fingerprint of $u$, where $d^u_i$ denotes the $i$th most visited web domain by this user. To analyze uniqueness given a fingerprint length $n$, we count the number of non-duplicate $m^n_u$ within our data, treating fingerprints as sets to disregard the order of elements. We use the Jackknife method~\cite{efron2021computer} to estimate the standard error of the uniqueness in our analyses (Fig.~\ref{fig1}a-b and Fig.~\ref{fig1}f-g).

To evaluate diversity in users' fingerprints, we compare each unique user against all others. For a fingerprint of length $n$ and for each unique user $u$, we first randomly select a domain $d^u_{r1}$ from $m^n_u$ and form a set $S(\{d^u_{r1}\})$ comprising users with $d^u_{r1}$ in their fingerprints. Next, we randomly choose another domain $d^u_{r2}$, without replacement, from $m^n_u$, and define a set $S(\{d^u_{r1}, d^u_{r2}\})$ for users having both $d^u_{r1}$ and $d^u_{r2}$ in their fingerprints. This selection process continues until $|S(\{d^u_{r1}, \cdots, d^u_{rl}\})|=1$, and the set only contains user $u$. We note that $l \leq n$ and represents the number of steps (or domains) necessary to distinguish user $u$ from other unique users; for each user, we find the average of $l$ over 300 repetitions of this procedure (Fig.~\ref{fig1}d). 

To assess the efficacy of fingerprints in re-identifying individuals, we partition the data into two sequential time slices. First, we select a specific number of hours, denoted as $t$, to create users' fingerprints. Each fingerprint is constructed using $t$ hours of browsing data for each user.  Next, we attempt to re-identify the same users in a subsequent time slice of equal duration, the next $t$ hours, using the fingerprints generated from the first time slice. In this analysis, we use fingerprint lengths $4$, $5$, $10$, and $15$ (Fig.~\ref{fig1}e).

\bibliographystyle{naturemag}
\bibliography{main}

% \section*{Acknowledgements (not compulsory)}
% Acknowledgements should be brief, and should not include thanks to anonymous referees and editors, or effusive comments. Grant or contribution numbers may be acknowledged.

\section*{Data Availability}
The sources of all empirical data used in our analyses are available at \url{https://doi.org/10.5281/zenodo.4757574}. 

% \section*{Code Availability}
% All relevant code used in this study will be available at \url{https://github.com/macoj/assortativity}. % and its respective Zenodo repository~\cite{zenodocode}.

\section*{Ethics declarations}
\subsection*{Competing interests}
 The authors declare no competing interests.
 
%  \section*{Authors Contributions}
% XXX proposed the project, wrote relevant code, carried out analytical analyses, wrote the first draft, and reviewed the final manuscript.

\section*{Acknowledgments} 
We would like to thank Respondi AG for providing the web tracking and survey data used in this paper at no cost for research purposes, with special thanks to Fran\c{c}ois Erner and Luc Kalaora at Respondi for their valuable insights and assistance with data extraction. 

\end{document}